\documentclass[11pt,dvips]{article}
\usepackage{epsfig,times} 
\usepackage{picinpar}
\usepackage{wrapfig}
\usepackage{floatflt}
\makeatletter
\renewcommand{\section}{\@startsection{section}{1}{0in}
        {0.4\baselineskip}{0.1\baselineskip}{\Large\bf}}
\renewcommand{\subsection}{\@startsection{subsection}{2}{0in}
        {0.25\baselineskip}{-\baselineskip}{\large\bf}}
\renewcommand{\subsubsection}{\@startsection{subsubsection}{3}{0in}
        {0.1\baselineskip}{-\baselineskip}{\normalsize\bf}}
\makeatother
\def\uka { \raisebox{-0.5ex} {\mbox{$\stackrel{<}{\scriptstyle \sim}$}}}

\newcommand{\zillions}{$4.5\cdot 10^8$\,}
\newcommand{\alex}{Alexandreas {\em et al.} 1993}

\newcommand{\elman}{Goetting {\em et al.} 1999}
\newcommand{\bird}{Bird {\em et al.} 1995}
\newcommand{\hayas}{Hayashida {\em et al.} 1996}
\newcommand{\horns}{Horns~1999}
\newcommand{\hunter}{Hunter {\em et al.} 1997}
\newcommand{\lima}{Li and Ma 1983}

\newcommand{\taylor}{Taylor {\em et al.} 1987}
\newcommand{\wapap}{Krawczynski {\em et al.} 1996}
\newcommand{\schmele}{Schmele 1998}

\begin{document}

%
\thispagestyle{myheadings}
%
\markright{OG 3.2.24}
\begin{center}
%
{\LARGE \bf Anisotropies of Cosmic Rays and Search for Intergalactic Cascades from the direction of the Highest Energetic Cosmic Ray Events with the HEGRA Scintillator Array}
\end{center}

\begin{center}
%
%
{\bf D.\,Horns and D. Schmele for the HEGRA Collaboration}\\
{\it II. Inst. f. Exp.Physik, Universit\"at Hamburg, D-22761 Hamburg, Germany\\
}
\end{center}

\begin{center}
{\large \bf Abstract\\}
\end{center}
\vspace{-0.5ex}
 Data taken with the HEGRA Scintillator Array within 2.5  years have been used to
search for spatial anisotropies in the arrival directions of cosmic rays
(E $>$ 20 TeV).  For this purpose partial sky survey maps are produced, carefully
taking the detector response and changing conditions in the atmosphere into account.
 In this paper, results on the search for TeV $\gamma$--emission correlated
with the Galactic Plane 
and the Gould Belt, a local Galactic region with an enhanced concentration of young bright stars,
molecular and dust clouds, are presented. 
The upper limit for the emission from the Galactic disc of
$\Phi_\gamma/\Phi_{CR} \approx 10^{-4}$ for $|b_g|\le5^\circ$ ($b_g$ denoting the Galactic latitude) and $E>42$ TeV 
 imposes constraints on the extrapolation of the $\gamma$-ray flux measured by EGRET. 
Furthermore, a search for extended TeV $\gamma$--emission correlated with the arrival directions of the
cosmic rays with energies well beyond the Greisen-Zatsepin-Kuzmin cutoff has been
performed. For this purpose, the directions of 10 events with energies beyond $5\cdot 10^{19}$ eV including the
events belonging to three AGASA clusters
 and the most energetic event ($E=320$ EeV)
 detected by the Fly's eye detector 
have been examined for correlated TeV emission. Such a correlation can be expected from sufficiently high energetic nuclei 
inducing electromagnetic cascades on the diffuse intergalactic (2.7 K) background radiation, eventually
producing TeV photons.
Interestingly, the direction of the most energetic event
yields the highest excess with a chance probability of 1.8\%.

\vspace{1ex}

%
%

\section{Introduction}
\label{intro.sec}
 Charged cosmic rays of $E$\uka $10^{16}$ eV
 are repeatedly deflected in the $\approx\mu$G magnetic field
of the Galaxy causing a diffusive transport.
An observer within the Galaxy is expected to observe essentially an isotropic distribution
of arrival directions. 
Anisotropies of small amplitude in the arrival directions of cosmic ray on earth are expected
due to the earth movement relative to an isotropic cosmic ray flux (Compton--Getting effect) or
possible nearby sources or local magnetic fields. The search for a siderial dipolmoment has been carried out
using the same data set but is not presented here.

Furthermore, and this is the topic of this
paper, anisotropies in the arrival directions of air showers may be caused by photons. These photons may be produced in collisions 
of charged cosmic rays with nuclei of the interstellar medium, producing  $\pi^0$ which subsequently decay into energetic photons. The flux of photons from $\pi^0$
production is proportional to the density of cosmic rays and of interstellar matter, which is strongly concentrated towards the Galactic disc. The EGRET experiment on board the Compton Gamma Ray
 Observatory has measured gamma-ray emission from the Galactic disk in the
 energy range from 100 MeV to 50 GeV (\hunter). For higher energies, no diffuse emission has been
 detected so far.

 Another possible production mechanism of TeV photons is associated with the
 propagation of ultra high energy cosmic rays (UHECR) in the Cosmic Microwave Background (CMB)
 radiation.   An UHECR proton beyond the
 GZK cutoff ($E>5\cdot 10^{19}$eV) 
 propagating through the extragalactic space suffers energy loss in inelastic
 collisions with photons of the CMB radiation.
 A variety of inelastic processes channel energy into electromagnetic cascades driven by repeated Compton scattering
 and pair production processes with the CMB radiation. 
Once the energy of the photons drops below the pair production
 threshold energy, no more absorption occurs. As a possible consequence any sufficiently
 high energetic particle is accompanied by a large number of photons with
 energies below $\approx 10^{14}$eV. In case
 of a steady state injection mechanism and a low ($<10^{-9}$G) extragalactic magnetic field, the cascade photons would be
 numerous and the arrival direction would be well associated with the
 direction of the UHECR. 
 
 \section{Experimental setup and data selection}
 \label{setup.sec}
 The HEGRA Scintillator Array is part of the multi-component Air Shower 
 Detector located on the Canary Island of La Palma ($28.8^\circ$N $17.9^\circ$W, $2200 $m a.s.l.). 
 The angular resolution is estimated to be less than 1$^\circ$ and the threshold for 
 air showers of vertical incidence is  20 TeV for photons.  The energy threshold for photon induced showers increases to
 $\approx 1$ PeV for showers with a zenith angle of $60^\circ$. 
 Details on the performance of the scintillator array 
 can be found in \wapap. 
 
 The data has been selected carefully from the data taking period November 1993 -- June 1996 by
monitoring all available properties of the detector performance. Altogether \zillions showers (constituting 80\% 
of the data taken in this time period)  have been used for this analysis. 

The better accuracy in angular resolution 
obtained with the wide angle Cherenkov Counter array AIROBICC does not improve the sensitivity for
 extended source regions and for the benefit of a homogenous detector setup, data taken simultaneously with AIROBICC have been excluded from this analysis. A partial sky survey for pointsources with AIROBICC data will be presented in a separate contribution (\elman). 

\section{Data Analysis}
\label{analyses.sec}

A crucial task in the search for large scale anisotropies is to determine an experimental expectation for
an isotropic distribution of cosmic ray arrival directions to be used as background. 
This Background Map is  to be compared with the actual distribution of cosmic ray arrival directions.
Since there is no separate and independent measurement of the background available, one has to rely upon methods
modeling the background from the data itself.  

Conventional methods (see e.g. \alex) could not be applied because of the fact that the
time dependence in the detection rate $R=R(t,\theta,\phi)$ ($\theta ,\phi$ denoting local coordinates) can not be
factorized for the HEGRA data. Instead, it was found to be necessary to describe the rate by the following 
function (\schmele):
$$R=R(t,\theta,\phi)=R(\theta,\phi)\cdot\exp\left(\frac{P_0-P(t)}{\Lambda\cdot\cos(\theta)}\right)$$
The values of $\Lambda=105.4$ g/cm$^2$ and $P_0=774.0$ g/cm$^2$ were determined
from the data. $P(t)$ denotes the measured air pressure at a given time.

 By this empirical function, the 2-dimensional shape of the detection rate
 $R(\theta,\phi)$ 
is modified depending upon the air
pressure and the zenith angle. The method to determine a Background Map in celestial coordinates can be split roughly into three steps:
\begin{enumerate}
\item {\bf Sensitivity Map:} Fill all events in a map of local coordinates: $R(\theta,\phi)$ 
\item {\bf Measurement Condition Map:} Fill all events in a 2-dimensional map of
  siderial daytime and air pressure ($t_{sid},P$).
\item {\bf Background Map:} Calculate for {\em each bin} of the Measurement
  Condition Map a weighted Sensitivity Map using the overall Sensitivity Map
  modified by the empirical correction given above and transform this map in celestial
  coordinates.  
\end{enumerate}

 The weights used in the last step are taken from the number of entries in
each bin of the Measurement Condition Map. The normalization is chosen to
ensure that the sum over all bins of the Background Map is equal to the sum
over all bins in the Data Map.
Since the
observation period spans 2.5 years, all contributions by possible sources will
be smeared out in the Background Map. The Data and the Background Map contain square bins of $0.1^\circ \times 0.1^\circ$  in celestial coordinates.

 On the basis of the Data and the Background Map, searches can be performed on
 different angular scales. For this purpose, the search bin size has been varied from $1^\circ$
 to $10^\circ$. The smallest bin corresponds to a search for pointsources,
 since the angular resolution of the detector is approximately
 $1^\circ$. To obtain a probability for a given deviation from the background,
 the number of events within the search bin are summed up
 separately for the Data and the Background Map. A significance is calculated
 according to \lima. 
\section{Results on Anisotropies}
\label{resultsa.sec}
 
We report in this paper on the results from searches for emission from the Galactic
  plane and the Gould Belt. The results of searches for 
extended emission from other large scale structures can be found in
 (\schmele). 
\subsection{Galactic Plane}
 So far, the modeling of the EGRET data does not
 account completely for the flux above 100 MeV (\hunter). The search for emission at even higher energies (above 20
 TeV with the HEGRA detector) could help constrain the underlying production
 mechanisms. The search bin size has been chosen to be $10^\circ$. This is in
 accordance with the emission region measured by EGRET.
 The upper limit on the relative content of emission from the Galactic plane
 for \mbox{$|b_g| \le 5^\circ$}, \mbox{$0^\circ \le l_g \le 255^\circ$} ($b_g$ and $l_g$ denoting
galactic latitude and longitude respectively) and \mbox{$E>42$ TeV} is
 \mbox{$\Phi_\gamma/\Phi_{CR} < 1.6 \cdot 10^{-4}$.} This upper limit is about
 a factor two above the predicted $\gamma$-flux due to nucleon-nucleon interactions.
 Compared with the extrapolation of the EGRET flux at the
 considered part of the Galactic disk, the upper limit  constrains the integral spectral index for
 energies above 50 GeV to be steeper than 1.3 (under the assumption of a pure
 power law spectrum).

\subsection{Gould Belt}  
The population of young stars (spectral class O and B) is concentrated towards
the Galactic plane. The subgroup of stars that are within one 1kpc of the
solar system is aligned along a plane inclined by $18^\circ$ against the
Galactic plane, named {\em Gould Belt}, containing also several molecular
and dust clouds (\taylor). Again the search bin size has been chosen to be
$10^\circ$. The significance for the deviation from an isotropic background 
in this plane is 3.1 $\sigma$.
An unconstrained search for the
equatorial plane with the highest significance is given 
$$b_X=0^\circ + (30\pm 5)^\circ\cdot\sin(l-(340\pm5)^\circ)$$
yielding 5.9$\sigma$. So far, to our knowledge no motivation for an emission
from this plane exists.

\section{Search for Intergalactic Cascades}
\label{cascades.sec}
The energy dissipated by UHECR beyond the GZK cutoff in inelastic scattering
is partially channeled into
electromagnetic cascades driven by inverse Compton scattering and repeated
pair production. Once the photons produced are below the threshold for pair
production with the CMB photons 
they are observable at energies below $\approx 100$ TeV. The lower
energy bound of observable photons depends crucially on the magnitude of the magnetic field
strengths. Synchrotron energy loss is a concurrent energy loss mechanism to
inverse Compton scattering and scattering on magnetic fields could weaken a
directional correlation with decreasing energy. Detailed simulations of the
 cascading process are presented 
elsewhere in these proceedings (\horns).
 We have searched for directional correlations with the most energetic event detected by the
Fly's Eye Group (\bird) with an energy of 3.2 $\cdot 10^{20}$ eV (in the following FE320)
 and the
 events belonging to three
 of the so called AGASA Clusters (\hayas). The position yielding
 the most significant excess is the direction of the FE320 event. 
\begin{figwindow}[2,r,{%
\epsfig{file=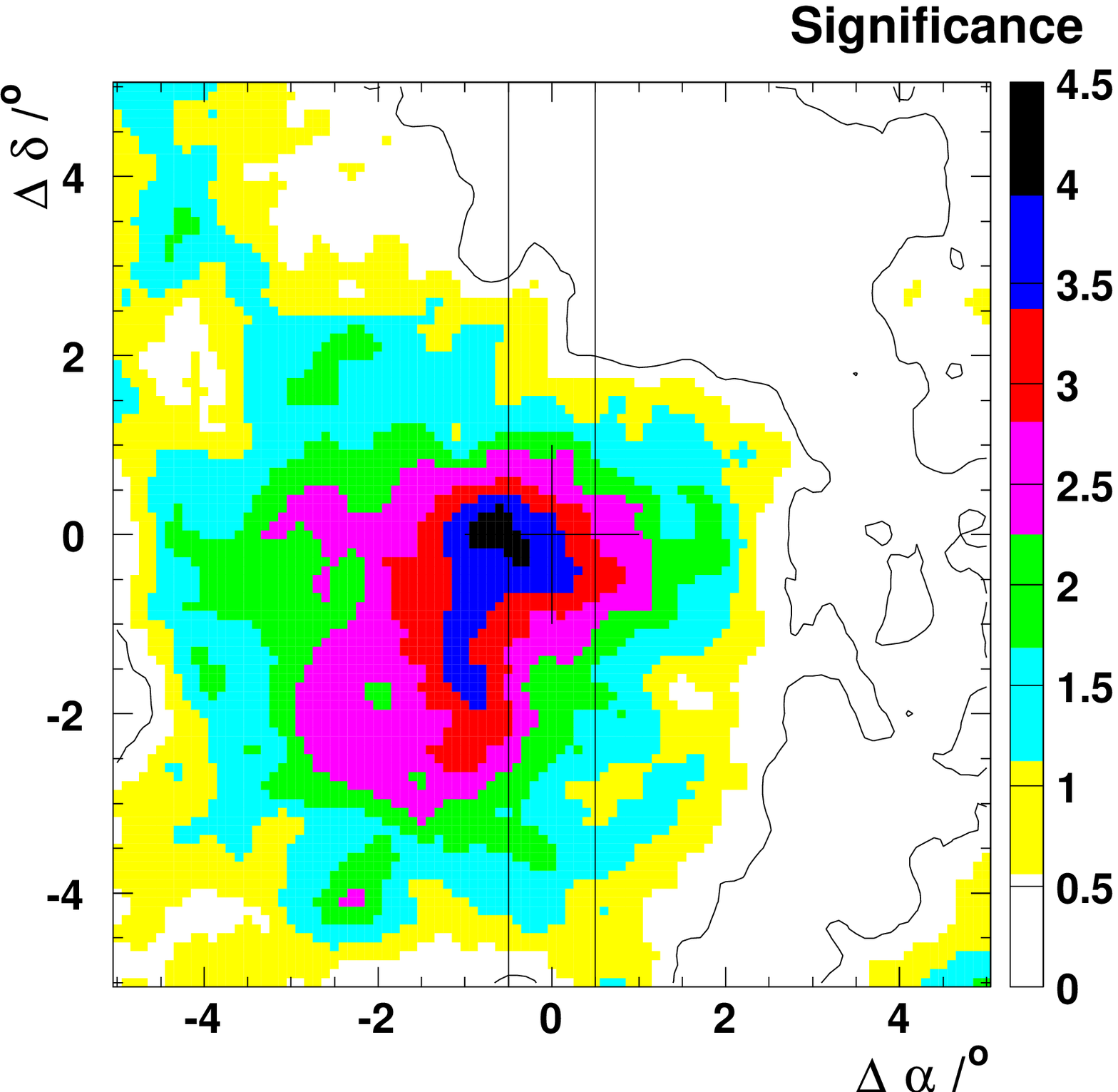, width =8.0cm}},
{The 2-dimensional map displays the positive significances for test positions
  in the vicinity of the FE320 direction marked by a cross (the rectangular
  region gives the positional error on the direction as estimated in \bird)}]
 Since the directional correlation of the secondary particles can be weakened by the
 presence of magnetic fields, we have inspected a $30^\circ\cdot30^\circ$
 Field of View centered upon the FE320 direction and have searched all
 positions on a $0.1^\circ$ spaced grid with different search bin radii, ranging
 from $1^\circ$ to $5^\circ$ incremented by $0.25^\circ$ steps. The largest
 excess was found at a position shifted by $1^\circ$ in R.A. from the
 reconstructed FE320 direction and with a search bin radius of
 $2.75^\circ$. In Figure 1 is a part of the search region displayed, also
 showing the directional uncertainty of this individual event as given in
 \bird.  Using M$^{\mbox{te}}$ Carlo generated maps and repeating the search, a chance
 probability of finding such an excess with a search bin radius $>2.75^\circ$
 has been estimated to be less than 1.8\%. 

 Under the hypothesis of a signal correlated with the source of the FE320
 event being an extragalactic steady-state proton accelerator, it is
 interesting to note, that the ratio of the energy flux of the cascade photons
 and the energy flux associated with the FE320 event yields a
 source distance of less than 28 Mpc. 

\end{figwindow}

\section{Conclusions}
\label{conclusions.sec}
 We have reported on results obtained with 2.5 years of data taken with the
 HEGRA Scintillator Array. Selected results on searches for TeV emission correlated with the
 Galactic Disk and the Gould Belt have been
 presented. The upper limit on relative abundance of photons arriving from the
 Galactic Plane $\Phi_\gamma/\Phi_{CR}\le 1.6\cdot 10^{-4}$ 
$(E\ge 42 \mbox{TeV},\,\, |b_g| \le 5^\circ)$ restricts the extrapolation of the EGRET
  integral Flux measured at GeV Energies to be steeper than a power law with an
 index of $1.3$. 

The projection along
 the Gould Belt yields a significance of 3.1 $\sigma$. The projection with the
 maximum significance is not associated with any known enhanced matter density
 or prominent emission feature in other wavelengths.

 The search for correlations with possibly extragalactic cascades initiated
 by 10 selected UHECR events yielded the highest significance in association with
 the direction of the FE320 event. The chance probability of finding such an excess
 has been estimated to be $<1.8\%$.  Further studies with the HEGRA  Cherenkov
 Telescope System are underway.

\section{Acknowledgements}
 This work has been supported by the BMBF. We are grateful to the IAC and the ORM for
excellent working conditions. The HEGRA experiment is supported by the BMBF, DFG and
the CICYT.
%
%
%
%
%
%
\vspace{1ex}
\begin{center}
{\Large\bf References}
\end{center}
%
%
%
%
%
Alexandreas, D.\,E.\, {\em et al.} 1993, Nucl. Instr. Meth. {\bf 328}, 325. \\[-4pt]
Bhattarcharjee, P.\, and Sigl, G.\, {\tt astro-ph/9811011}, submitted to Phys.
Rep.\\[-4pt]
Bird, D.\,J.\, {\em et al.} 1995, ApJ {\bf 441}, 144.\\[-4pt]
Cronin, J.\,W.\, {\em et al.} 1992, Nucl. Phys. B {\bf 28B}, 213.\\[-4pt]
Hayashida, N. {\em et al.} 1996, Phys. Rev. Lett. {\bf 77}, 1000.\\[-4pt]
Goetting, N. {\em et al.}  OG 2.1.33 Proc. 26th ICRC (Salt Lake City, 1999).\\[-4pt]
Horns, D.  OG 3.2.37  Proc. 26th ICRC (Salt Lake City, 1999).\\[-4pt]
Hunter, S.\,D.\, {\em et al.} 1997, ApJ {\bf 481}, 205. \\[-4pt]
Krawczynski, H.\, {\em et al.} 1996, Nucl. Instr. Meth {\bf 383}, 431. \\[-4pt]
Li, T.P. and Ma, Y.\,Q.\, 1983, ApJ {\bf 272}, 317. \\[-4pt]
Schmele, D.\, 1998, PhD Thesis, Univ. Hamburg.\\[-4pt]
Taylor, D.\,K.\, {\em et al.} 1987, ApJ {\bf 315}, 104.
\end{document}